\documentclass[11pt,a4paper,twoside]{article}
\usepackage{fancyhdr}
\usepackage{enumerate,xcolor}

\usepackage[top=100pt,bottom=110pt,left=100pt,right=100pt]{geometry}

\usepackage{float}
\usepackage{caption}
\usepackage{cite}

\usepackage{amsmath}
\usepackage{amsthm}
\usepackage{amsfonts}
\usepackage{amssymb}
\usepackage{mathrsfs}
\usepackage{graphicx}
\usepackage{subcaption}
\numberwithin{equation}{section}

\theoremstyle{remark}

\theoremstyle{definition}

\title{A qualitative mathematical model \\of the immune response under the effect of stress}
\author{Maria Elena Gonzalez Herrero and Christian Kuehn}
\begin{document}

\maketitle

\begin{abstract} 
    In the last decades, the interest to understand the connection between brain and body has grown notably. For example, in psychoneuroimmunology many studies associate stress, arising from many different sources and situations, to changes in the immune system from the medical or immunological point of view as well as from the biochemical one. In this paper we identify important behaviours of this interplay between the immune system and stress from medical studies and seek to represent them qualitatively in a paradigmatic, yet simple, mathematical model. To that end we develop a differential equation model with two equations for infection level and immune system, which integrates the effects of stress as an additional parameter. We are able to reproduce a stable healthy state for little stress, an oscillatory state between healthy and infected states for high stress, and a ``burn-out'' or stable sick state for extremely high stress. The mechanism between the different dynamics is controlled by two saddle-node in cycle (SNIC) bifurcations. Furthermore, our model is able to capture an induced infection upon dropping from moderate to low stress, and it predicts increasing infection periods upon increasing before eventually reaching a burn-out state. 
\end{abstract}


\let\thefootnote\relax
\footnotetext{Maria Elena Gonzalez Herrero \\ Department of Mathematics, Technical University Munich, 85748 Garching, Germany \\ Email: me.gonzalez@tum.de\\}
\footnotetext{Christian Kuehn \\ Department of Mathematics, Technical University Munich, 85748 Garching, Germany  \\ External Faculty, Complexity Science Hub Vienna, Josefst\"adter Str. 39, 1080 Vienna, Austria \\ Email: ckuehn@ma.tum.de}

\section{Background}

    Although it is common knowledge that stress can hurt our health, scientific results regarding this are fairly recent. The first studies attempting to show this connection between the psychological and the physical state were developed in the first half of the 20th century \cite{Cannon32, Selye36}. In the second half of the century the corresponding field of psychoneuroimmunology (PNI) established the interplay of the central nervous system (CNS), the endocrine system, in other words the hormonal messenger system of the body, and the immune system \cite{Solomon64, Holmes67, Ader01, Glaser05}. The first goal of this field is to specify a good definition of stress to be able to work with and thus also a way to categorize it \cite{Elliot82}. The most common way to do this is mainly by the duration, e.g., one may contrast very brief public speaking to extremely long-time care of a spouse \cite{Segerstrom04}. Furthermore, it is important to quantify stress factors, e.g. how and to what extend we can see sports as positive or negative stressor in the same way as psychological ones \cite{Hoffman94}. Further, PNI wants to understand the biochemistry of how behavioral and psychological effects, in particular stress, are reflected in our bodies, i.e. how stress ``gets inside the body'' \cite{Segerstrom04}. Although there is still much to learn, we have now a basic understanding of the hormonal reactions of our body to stress and how they affect the immune system, e.g. by specific receptors on cells responsible for the immunity \cite{Glaser05}. In addition, there are several studies regarding immunological and medical consequences of stress~\cite{Kiecolt85, Andersen04}, where a key goal is to understand and counteract negative consequences. In this context, the difficulties of extensive clinical trials and the lack of a commonly accepted definition, categorisation or measurements of stressors are most evident. This is why Segerstrom and Miller performed in 2004 a meta-analytical study of numerous results going back 30 years \cite{Segerstrom04}. They showed that in general stress has a negative effect on the immune system by either decreasing the number of e.g. killer cells or simply disturbing the equilibrium between the different components. Nevertheless a brief stress, like public speaking, can also have a positive effect and even longer short-term stressors like examination periods for students can have a beneficial effect during but cause decrease immunity afterwards \cite{Dorian85, Burns02}.

\section{The model}

    Having some fundamental understanding of how our immune system reacts to stress, we want to find a \emph{simple mathematical model} of differential equations which is able to represent the \emph{basic qualitative interaction} between an infection and our \emph{immune system} taking into account the \emph{effect of stress}.\\
    
    Since we are only interested in a qualitative analysis of the system we are going to work with normalised dynamical variables $x$, representing the level of activity of the immune system, and $y$, as the level of infection or sickness, with values ranging between 0 and 1. The stress is included by a parameter $s$ also normalised between 0 and 1 where 0 corresponds to no stress at all and 1 is the maximal level of stress limited by the assumption that a person cannot feel infinitely strong stress. 
    
    %Since, for simplicity, the stress in our model is fixed we are not able to reflect a one to one correlation of stressors of different durations. Nevertheless we could consider a prolonged stress as a ``stronger'' stress to still obtain some comparison.
    
    To specify the dynamics of the infection we are going to built upon principles of mathematical biology~\cite{Murray1,Mueller15}, where in multiple sub-disciplines such as ecology, neuroscience, or biomechanics, one aims to generic polynomial nonlinearities to represent basic effects. Based upon population dynamics, we consider logistic growth and Allee effect for the immune system intrinsic dynamics. On this regard, our first assumption is that, as long as our immune system has some activity, the infection should have two stable states, the ``sick''-state and the ``healthy''-state. Furthermore, since we are constantly coming in contact with different viruses and bacteria, we are interested in having a slightly raised base line instead of the ``healthy''-state being $y=0$. Our second assumption is that, if there is no immune system or if it is too weak then the infection should spread up to its maximal capacity. Combining everything we arrive at the equation
    \begin{equation}
        y'= -xy + r\sqrt{1-y}\left(q(y-y_1)(y-y_0)+(1-q)y\right)
    \end{equation}
    where we used a square root instead of the standard multiplicative term $(1-y)$ from population models. Finally we choose the parameter values $r=2$, $y_0=y_1=0.1$ and $q=0.95$.\\
    
    In contrast to the previous derivation of the dynamics for $y$, our approach for finding an appropriate equation for the immune system is based purely on the geometry necessary to capture the effects we aim to model. We start with the case without stress. It is clear that without stress we should have a unique stable equilibrium with $0\ll x<1$ and $0<y\ll 1$ corresponding to a normal healthy state. Since too much stress generally induces an infection, this fixed point should only exist for $s<s_0$ with some threshold level of stress $0<s_0<1$.\\
    
    Furthermore, we also want to reflect the fact that, while a moderate stress can make slightly more resistant to infections, at the same time, its sudden drop to normal levels can be detrimental increasing the likelihood of infections, e.g., this naturally occurs for short-term stressors. To capture this effect we want to have a second unstable equilibrium with $0<y\ll 1$. As $s$ increases the $x$-coordinates of both fixed points should slowly start decreasing and approaching each other such that for some $0<s_1<s_0$ the stable equilibrium crosses the original position of the unstable one. If we would now drop the stress back to $s=0$ it would in fact trigger the expected infection before converging to the healthy state.\\
    
    Finally, it is reasonable to assume that an extreme (prolonged) stress can prevent the body from recovering from an infection, giving a ``burn-out'' state, such that for $s$ close to 1 we want to have a fixed point where $0\ll y<1$.\\
    
    Again combining all the assumptions above, one simple polynomial nonlinear differential equation capturing these effects is given by
    \begin{equation}
    \label{eq:xeq}
        \begin{aligned}
            x' = & ~~y - 500\left(\frac{1}{5}(kx)^5 - \frac{1}{4}(3x_0+x_1)(kx)^4 + (x_0^2+x_0x_1)(kx)^3\right.\\
            & ~~\left.-\frac{1}{2}(x_0^3+3x_0^2x_1)(kx)^2+x_0^3x_1(kx)\right)+0.2
        \end{aligned}
    \end{equation}
    presenting an S-shaped nullcline with $x_0=0.3$ and the parameter $k=1.3-0.3(1-s)^4$ to control the shift to the right of the equilibria and $x_1=0.8-0.2s^2$ responsible for the disappearing of the stable healthy state. Of course, the actual numerical values in~\eqref{eq:xeq} are not the key aspect but the geometry of the dynamics and the qualitative description of the different aspects of immune reaction, infection level, and stress. 
    
\begin{figure}[h]
        \hspace*{-3mm}\includegraphics[width=0.5\textwidth]{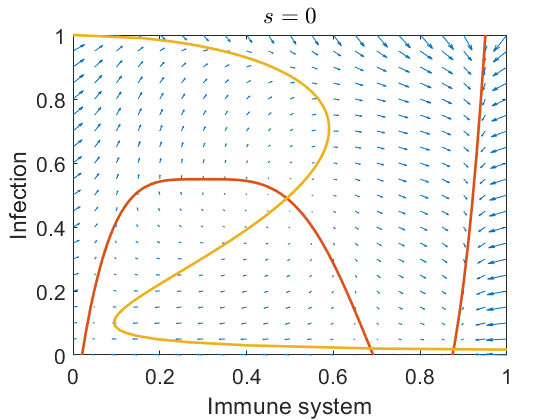}
        \hspace*{3mm}\includegraphics[width=0.5\textwidth]{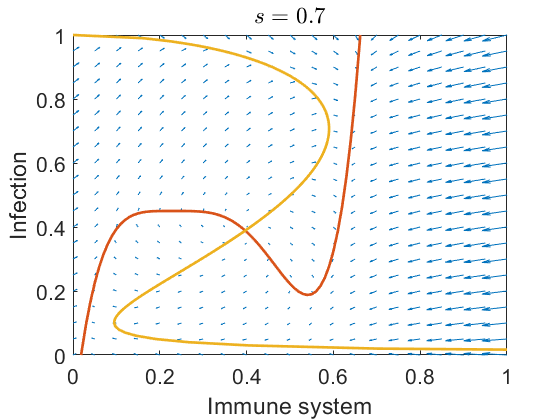}\\
        \includegraphics[width=0.5\textwidth]{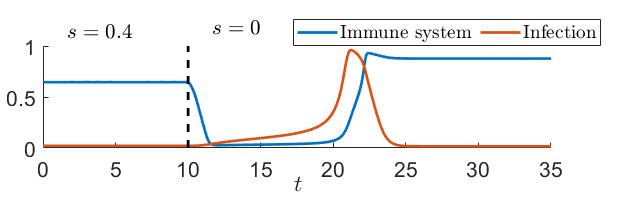}
        \hspace*{2mm}\includegraphics[width=0.5\textwidth]{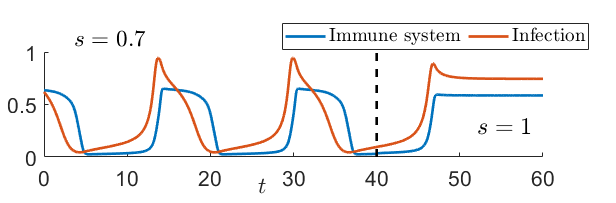}\\
        \hspace*{-3mm}\includegraphics[width=0.5\textwidth]{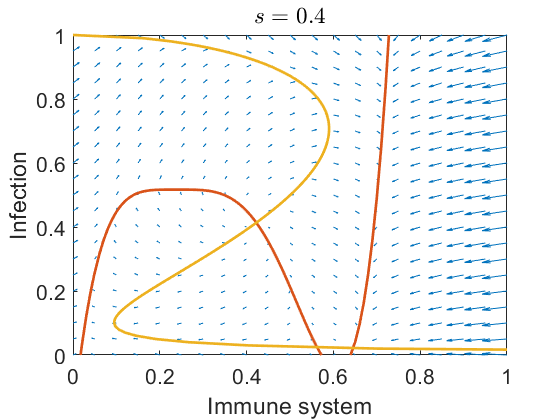}
        \hspace*{3mm}\includegraphics[width=0.5\textwidth]{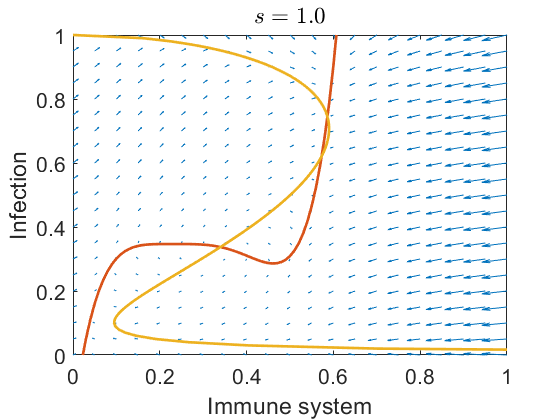}
        \caption{Phase portraits of system \eqref{eq:main} and its $x$-nullcline (orange) and $y$-nullcline (yellow) for different values of $s$ together with the corresponding time series. On the left the healthy state is stable and the time series shows the isolated sickness, when changing to no stress $s=0$ after a moderate stress $s=0.4$ at the dashed vertical line, i.e., switching from the top phase portrait to the bottom one. In the right column, the is no healthy equilibrium and we either oscillate for high stress $s=0.7$ or we stay sick (``burn-out'' state) for extreme stress level $s=1$. Again, a time series is shown, where we switched from the top to the bottom dynamical phase portrait at the dashed black vertical line.}\label{fig:phasePortrait}
    \end{figure}

    In summary, the complete system we arrive at is
    \begin{equation}\label{eq:main}
        \begin{aligned}
            x' = & ~~y - 500\left(\frac{1}{5}(kx)^5 - \frac{1}{4}(3x_0+x_1)(kx)^4 + (x_0^2+x_0x_1)(kx)^3\right.\\
            & ~~\left.-\frac{1}{2}(x_0^3+3x_0^2x_1)(kx)^2+x_0^3x_1(kx)\right)+0.2\\
            y' = & ~~\alpha\left(-xy + 2\sqrt{1-y}\left(q(y-y_1)^2+(1-q)y\right)\right)
        \end{aligned}
    \end{equation}
    with the parameter values $x_0=0.3$, $q=0.95$,  $y_1=0.1$ and
    $$k=1.3-0.3(1-s)^4~~~~\text{ and }~~~~ x_1=0.8-0.2s^2$$ 
    integrating the effect of stress into the equations. The additional parameter $\alpha$ controls the scale of change of the infection $y$ with respect to the immune system $x$. Figure \ref{fig:phasePortrait} shows the corresponding phase portraits and time series for multiple values of $s$. And indeed, our model presents different dynamic regimes upon different stress levels, which we can observe in the phase portraits and time series shown in Figure~\ref{fig:phasePortrait}.
    
\section{Bifurcation analysis}

\begin{figure}[h]
        \includegraphics[width=\textwidth]{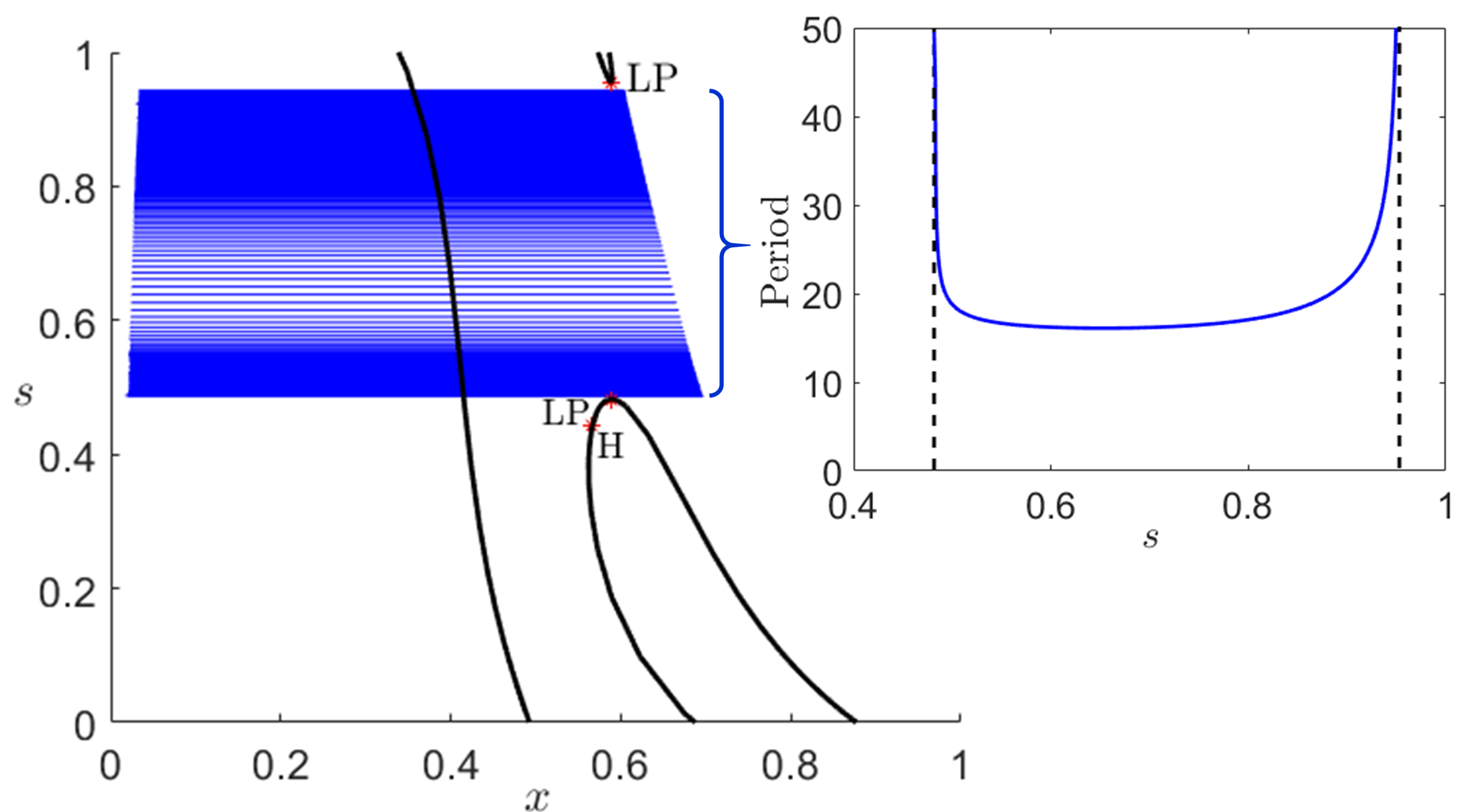}
        \caption{Bifurcation diagram (left panel) in the $(x,s)$-plane for $\alpha=2.5$. Black curves mark equilibrium points, while the blue region consists of limit cycles, including the period of the limit cycles for $0.48<s<0.95$ (right panel). $LP$ mark the limit points or saddle-node bifurcation points (in this case saddle-node in limit cycle (SNIC) bifurcation points) while $H$ marks a neutral saddle, where the eigenvalues $\lambda_{1,2}$ of the Jacobian at the equilibrium satisfy $\lambda_1=-\lambda_2$. The diagram has been computed using MatCont~\cite{Dhoogeetal}.}\label{fig:bifurcation}
    \end{figure}

     To analyse in more detail the change between the different dynamical regimes, we take a look at the corresponding bifurcation diagram~\cite{Kuznetsov} in Figure \ref{fig:bifurcation}. For $s=0$ we have 3 equilibria: an unstable spiral, a saddle and a stable node. As $s$ increases the stable node and the saddle collide and disappear in a saddle-node in cycle (SNIC) bifurcation for $s\approx 0.48$. At $s\approx 0.95$ we find a second SNIC, where another pair of a stable node and a saddle appear. In addition to the equilibria mentioned above, in the parameter regime between the two bifurcation points we find that all orbits except the unstable equilibrium have to converge to a limit cycle as $t\to\infty$. This can be easily shown using the Poincaré-Bendixson Theorem on the unit square. Note that SNIC bifurcations are of co-dimension one, i.e., they are generic/typical in planar systems with one parameter presenting a mechanism to obtain oscillations. They have been observed already in many models ranging from neuroscience \cite{Xie08, Maesschalck15} to mechanics \cite{Czaplewski18}. They are accompanied by a heteroclinic loop for parameter values $s<0.48$ (or $s>0.95$) since the unstable manifolds of the saddle converge to the stable node. As $s$ approaches the bifurcation the two equilibria collide, one of the heteroclinic connections disappears and the second one becomes a homoclinic orbit. By perturbing $s$ further this homoclinic gives rise to a family of stable limit cycles for $0.48<s<0.95$. 
   
\section{Discussion}

In this work, we have initialized the qualitative mathematical model development for the interaction between immune system levels, infection dynamics, and stress level. Based upon established principles of theoretical biology, we developed a simple, yet powerful, planar dynamical system to analyze the influence of stress as an external parameter. Using only elementary nonlinearities, out model can represent (I) a stable healthy equilibrium for low stress, (II) the effect of sudden stress level drop from moderate to low levels inducing a single infection, (III) periodic outbreaks of infections under high stress, (IV) the transition to a burn-out state at very high stress. The key transition mechanism we have identified is a saddle-node in cycle (SNIC) bifurcation. This bifurcation transition actually makes the model predictive, e.g., it predicts that if stress levels are close to a burn-out stage but still below, then the oscillatory sick periods become longer. Although this looks like a natural prediction, it is one that we did not anticipate to be able to extract from our system at all during the modelling process.

Having established the ability of conceptual mathematical models to contribute to the understanding of stress, we believe that further model development and connecting qualitative mathematical models to more detailed biophysical principles as well as to clinical trials, could forge a effective path mitigating, and probably even more importantly, predicting in advance, the positive and negative effects of stressors.

%\small

\medskip

\textbf{Acknowledgements:} MEGH work was partly supported by a Hans-Fischer Senior Fellowship of the Technical University of Munich Institute for Advanced Study (TUM-IAS). CK acknowledges partial support of the VolkswagenStiftung via a Lichtenberg Professorship. 

\newpage

  \newpage
  
\begin{thebibliography}{20}
    
    \bibitem{Cannon32} W. B. Cannon, 1932, \textit{The wisdom of the body}, Kegan Paul and Co., Ltd., London
    
    \bibitem{Selye36} H. Selye, \textit{A Syndrome produced by diverse nocuous agents}, Nature 138, 1936
    
    \bibitem{Solomon64} G. F. Solomon and R. H. Moos, 1964, \textit{Emotions, immunity and disease. A speculative theoretical integration}, Arch Gen Psychiatry 11(6)
    
    \bibitem{Holmes67} T. H. Holmes and R. H. Rahe, 1967, \textit{The social readjusment rating scale}, J. Psychosom. Res. 11, 213-218
    
    \bibitem{Ader01} R. Ader, 2001, \textit{Psychoneuroimmunology}, Current Directions in Psychological Science 10(3), 94-98, doi:10.1111/1467-8721.00124
    
    \bibitem{Glaser05} R. Glaser and J. K. Kiecolt-Glaser, 2005, \textit{Stress-induced immune dysfunction: implications for health}, Nat. Rev. Immunol. 5, 243–251, doi:10.1038/nri1571
    
    \bibitem{Elliot82} G. R. Elliot and C. Eisdorfer, 1982, \textit{Stress and human health: An analysis and implications of research. A study by the Institute of Medicine, National Academy of Sciences}, New York: Springer Publishing
	
	\bibitem{Segerstrom04} S. C. Segerstrom and G. E. Miller, 2004, \textit{Psychological stress and the human immune system: A meta-analytical study of 30 years of inquiry}, Psychol. Bull. 130(4), 601-630, doi:10.1037/0033-2909.130.4.601
	
	\bibitem{Hoffman94} L. Hoffman-Goetz and B. K. Pedersen, 1994, \textit{Exercise and the immune system: a model of the stress response?}, Immunology today 15(8)
	
	\bibitem{Kiecolt85}  J. K. Kiecolt-Glaser, R. Glaser, D. Williger, J. Stout, G. Messick, S. Sheppard, D. Ricker, S. C. Romisher, W. Briner, G. Bonnell and R. Donnerberg, 1985, \textit{Psychosocial enhancement of immunocompetence in a geriatric population}, Health Psychology 4(1), 25–41, doi:10.1037/0278-6133.4.1.25
	
	\bibitem{Andersen04} B. L. Andersen, W. B. Farrar, D. M. Golden-Kreutz, R. Glaser, C. F. Emery, T. R. Crespin, C. L. Shapiro and W. E. Carson, 2004, \textit{Psychological, behavioral, and immune changes after a psychological intervention: a clinical trial.}, J. Clin. Oncol. 22(17), 3570-3580, doi:10.1200/JCO.2004.06.030
	
	\bibitem{Dorian85} B. J. Dorian, P. E. Garfinkel, E. C. Keystone, R. Gorczynski, D. M. Gardner, P. Darby and A. Shore, 1985, \textit{Occupational stress and immunity}, Psychosomatic Medicine 47,77
	
	\bibitem{Burns02} V. E. Burns, M. Drayson, C. Ring, D. Carroll, 2002, \textit{Perceived Stress and Psychological Well-Being Are Associated With Antibody Status After Meningitis C Conjugate Vaccination}, Psychosomatic Medicine 64(6), 963-970
	
	\bibitem{Murray1} J.D. Murray, 2002, \textit{Mathematical Biology I: An Introduction}, Springer, Berlin.
   
	\bibitem{Mueller15} J. Mueller and C. Kuttler, 2015, \textit{Methods and Models in Mathematical Biology}, Springer, Berlin
	
	\bibitem{Kuznetsov} Yu.A. Kuznetsov, 2004, \textit{Elements of Applied Bifurcation Theory}, Springer, Berlin

	\bibitem{Xie08} Y. Xie, K. Aihara and Y. M. Kang, 2008, \textit{Change in types of neuronal excitability via bifurcation control}, Phys. Rev. E 77, 021917
	
	\bibitem{Maesschalck15} P. De Maesschalck and M. Wechselberger, 2015, \textit{Neural excitability and singular bifurcations}, Journal of Mathematical Neuroscience 5(16), doi:10.1186/s13408-015-0029-2
	
	\bibitem{Czaplewski18} D. A. Czaplewski, C. Chen, D. Lopez, O. Shoshani, A. M. Eriksson, S. Strachan and S. W. Shaw, 2018, \textit{Bifurcation generated mechanical frequency comb}, Phys. Rev. Lett. 121, 244302
   
   	\bibitem{Dhoogeetal} A. Dhooge and W. Govaerts and Yu.A. Kuznetsov and H.G.E. Meijer and B. Sautois, 2008, \textit{New features of the software {MatCont} for bifurcation analysis of dynamical systems}, Math. Comput. Model. Dyn. Syst., 14(2), 147--175
\end{thebibliography}
\end{document}